\documentclass[%
 reprint,
 amsmath,amssymb,
 aps,
twocolumn,prl]{revtex4}

\usepackage{graphicx}
\usepackage{dcolumn}
\usepackage{bm}
\usepackage{epstopdf}

\def \diff{\ensuremath{\mbox{d}}}
\begin{document}

\preprint{APS/123-QED}

\title{Probing turbulence intermittency via  Auto-Regressive Moving-Average  models }

\author{Davide Faranda}
 \email{davide.faranda@cea.fr}
\affiliation{%
Laboratoire SPHYNX, Service de Physique de l'Etat Condens\'e, DSM,
CEA Saclay, CNRS URA 2464, 91191 Gif-sur-Yvette, France
}
 
\author{B\'ereng\`ere Dubrulle, Fran{\c{c}}ois Daviaud}
\affiliation{%
Laboratoire SPHYNX, Service de Physique de l'Etat Condens\'e, DSM,
CEA Saclay, CNRS URA 2464, 91191 Gif-sur-Yvette, France
}

\author{Flavio Maria Emanuele Pons}
\affiliation{%
Department of Statistics, University of Bologna, 
Via delle Belle Arti 41, 40126 Bologna, Italy .}
             

\begin{abstract}
 We suggest a new approach to probing intermittency corrections to the Kolmogorov law in turbulent flows based on the Auto-Regressive Moving-Average  modeling of turbulent time series. We introduce a new  index $\Upsilon$ that measures the distance from a Kolmogorov-Obukhov model in the Auto-Regressive Moving-Average models space. Applying our analysis to Particle Image Velocimetry  and  Laser Doppler Velocimetry  measurements in a von K\'arm\'an swirling flow, we show that  $\Upsilon$ is proportional to the traditional intermittency correction computed from the structure function. Therefore it provides the same information, using   much shorter time series.   
We conclude  that $\Upsilon$    is a suitable index to reconstruct the spatial intermittency of the dissipation  in both numerical and experimental turbulent fields.
\end{abstract}

\pacs{Valid PACS appear here}
\maketitle


\paragraph{Introduction.}  
 One of the few exact results known for isotropic, homogeneous and mirror-symmetric turbulence is the  4/5 - law  derived by Kolmogorv in 1941. It links  the longitudinal velocity increments $\delta u_\ell=(u(x+\ell)-u(x))$ to the mean rate of energy dissipation $<\epsilon>$ via:
 \begin{equation}
 \langle \delta u_\ell^3\rangle=-\frac{4}{5}\langle \epsilon \rangle \ell,
 \label{fourfifth}
 \end{equation}
 where $\langle \rangle$ denotes averaging. This exact relation was then generalized by Kolmogorov\cite{kolmogorov1962refinement} as a scaling law $\delta u_\ell \equiv (\epsilon \ell)^{1/3}$, where $\equiv$ means \textit{has the same statistical properties}. Should  $\epsilon$ be a non stochastic constant, the scaling law would imply self-similar behavior for the structure functions of order $p$, $S_p(\ell)=\langle \delta u_\ell^p\rangle$, that would scale like:
 \begin{equation}
F_p(\ell)\sim \epsilon^{p/3} \ell^{p/3}.
 \label{structure}
 \end{equation}
 For $p=3$, we recover the 4/5 - law. For $p=2$, this equation predicts a second order structure function that varies like $\ell^{2/3}$. By Fourier transform, this can be shown to be equivalent to a one dimensional energy spectrum scaling with wavenumber $k$ as : $E(k)\sim k^{-5/3}$, also known as the Kolmogorov spectrum \cite{kolmogorov1941local,obukhov1941distribution}. Both the 4/5 - law and the Kolmogorov spectrum have been measured and checked in many natural and laboratory isotropic turbulent flows \cite{frisch1996turbulence}. More generally, eq. (\ref{structure}) predicts a linear law for the exponent of the structure functions $\zeta(p)=\diff \ln F_p(\ell)/\diff \ln \ell=p/3$. However, as pointed out by Landau and recognized by Kolmogorov \cite{kolmogorov1962refinement}, there is no reason to assume that $\epsilon$ is a constant over space and/or time, so that it should rather be viewed as a stochastic process, that depends upon the scale $\ell$ at which it is measured: $\epsilon\equiv \epsilon(\ell)$. In such a case, the correct scaling of the structure function is rather
  \begin{equation}
F_p(\ell)\sim \langle \epsilon(\ell)^{p/3}\rangle \ell^{p/3}.
 \label{structurerefined}
 \end{equation}
This modified law therefore predicts correction to the linear law $\zeta(p)=p/3$, that are connected to the intermittent nature of the dissipation. For example, a log-normal model for of the dissipation (a suggestion by Landau and Obukhov) implies a quadratic correction for the $\zeta(p)$. Other models have been suggested and lead to different corrections \cite{SL,dubrulle,benzi1984multifractal}. Intermittency corrections up to $p=4$ have now been measured in a variety of experimental and numerical flows and appear to be robustly consistent from an experiment to another (see e.g. the review of \cite{arneodo1996structure}). Corrections for larger values of $p$ are subject to   resolution and statistical convergence issues: the larger the scaling exponent, the larger the statistical sampling must be in order to capture the rare events. There is therefore presently no general consensus about the behavior of intermittency corrections at large order. This hinders progress in the understanding of the statistical properties of the energy dissipation.\\
In this Letter, we suggest a new approach to probing intermittency corrections based on the Auto-Regressive Moving-Average (ARMA) modeling of turbulent time series. We introduce a new index $\Upsilon$ that measures the distance from a Kolmogorov-Obukhov model in the ARMA space. Applying our analysis to velocity measurements in a von K\'arm\'an swirling flow, we show that this  index is proportional to the traditional intermittency correction computed from the structure function and therefore provides the same information, using   much shorter time series.

 \paragraph{Intermittency paramaters.}  
In most laboratory turbulent flows, available datasets are time series of values of a physical observable at a fixed point or obtained by tracking Lagrangian particles. This motivated the shift of paradigm from \textit{space} velocity increments to \textit{time} velocity increments defined as $\delta u_\tau=u(t+\tau)-u(t)$ and motivated measurements of the  time  structure function $G_p(\tau)=\langle (\delta u_\tau)^p\rangle$ and its local exponent $\chi_p=\diff \ln G_p(\tau)/\diff \ln \tau$. Of course, in situations where measurements are made on the background of a strong mean velocity $U$, scale velocity increments and time velocity increments can be directly related through the Taylor hypothesis $\ell=U\tau$. In situations such that the fluctuations are of the same order than the mean flow, however, the Taylor hypothesis fails. A suggestion has been made by \cite{pinton} to then resort to a \textit{local Taylor Hypothesis}, in which $\ell=\int dt u(t)$ where $u$ is the local rms velocity. This is equivalent to consider a scale such that $\ell\sim \tau  \delta u_\tau$  and may be seen as equivalent to modifying the \textit{space} Kolmogorov refined hypothesis into a \textit{time} hypothesis $\delta u_\tau\equiv (\epsilon \tau)^{1/2}$, that leads to:
\begin{equation}
 G_p(\tau)\sim \langle \epsilon^{p/2} \rangle \tau^{p/2}.
 \label{structurerefinedtime}
 \end{equation}
 Such scaling is   equivalent to the scaling obtained using the Lagrangian structure function.  In any case, we may define the intermittency as the deviation of the local exponents $\zeta^*_p=\zeta_p$ (space increments)  or $\zeta_p^*=\chi_p$ (time increments)  with respect to a linear behavior and may be quantified to first order by the parameter:
 \begin{equation}
 \mu=\zeta^*_2-\frac{2}{3}\zeta^*_3.
 \label{interm1}
 \end{equation}
 This factor is for example proportional to the log of the $\beta$ parameter of the log-Poisson model \cite{SL,dubrulle}, or to the $\mu$ parameter of the log-normal model \cite{kolmogorov1962refinement}. Note that it is also valid when the scaling exponents have been computed using the Extended Self-Similarity (ESS) \cite{benzi1993extended}, which is especially interesting in situations where turbulence is inhomogeneous and when the Taylor hypothesis does not hold. In the sequel, we  compare this intermittency index with another one, built in a purely statistical framework. \\
\begin{figure}
\includegraphics[width=.38\textwidth]{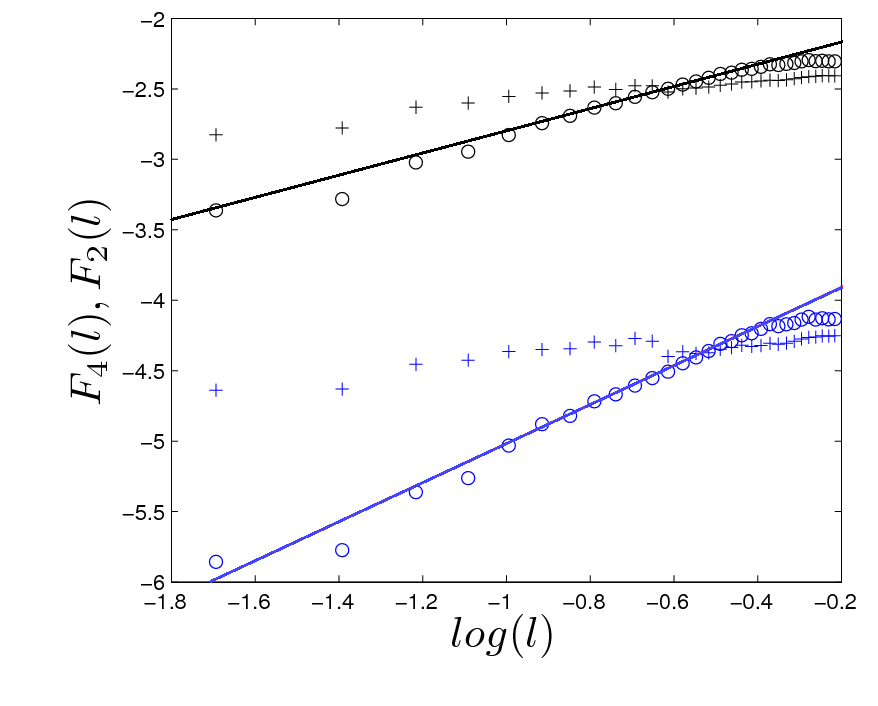}
 \includegraphics[width=.38\textwidth]{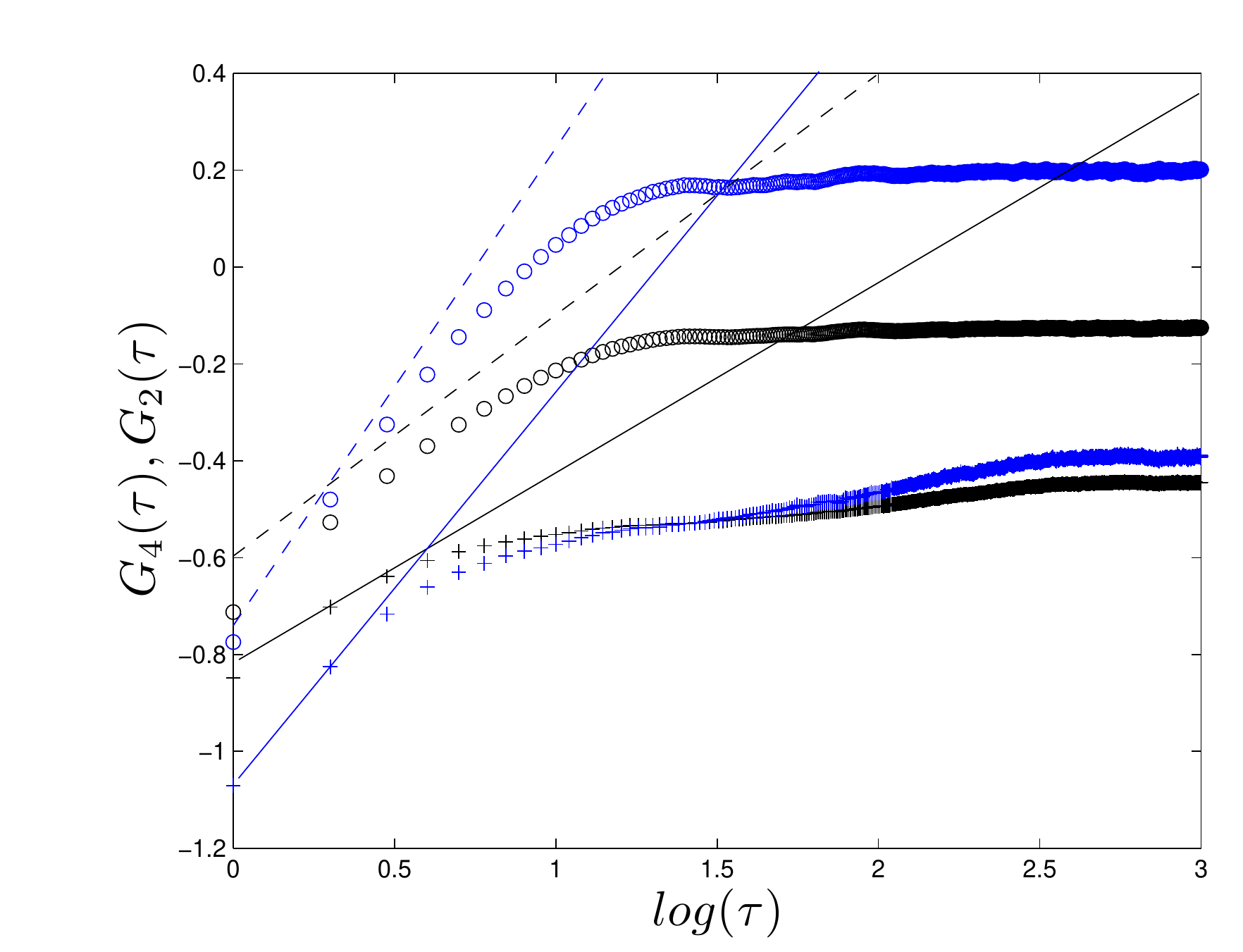}
\caption{Upper panel: space structure function $F_p(\ell)$  of order 2 (black) and 4 (blue) for two  points of the PIV grid. Circles: $R=0.16$, $Z=-0.04$, crosses: $R=0.77$, $Z=0.64$.  Lower panel: time structure function $G_p(\tau)$ of order 2 (black) and 4 (blue) for two  points of the LDV grid. Circles: $R=0$, $Z=0.35$, crosses: $R=0.52$, $Z=0.61$.  In the 2 panels lines are the Kolmogorov predictions: solid: Eulerian (black: $\ell^{2/3}$ or $\tau^{2/3}$ , blue: $\ell^{4/3}$ or $\tau^{4/3}$ ); dotted: Lagrangian (black: $\tau$, blue: $\tau^2$).} 
\label{g3l}
\end{figure}

Indeed, Thomson \cite{thomson-jfm-1987} showed that, in the Lagrangian framework, the "time" refined Kolmogorov hypothesis is in fact equivalent to a stochastic description in terms of an  Ornstein-Uhlenbeck process with suitable drift and noise term:
\begin{equation}\label{OU}
\diff u = -\frac{u}{T} \diff t + 	\sqrt{C_0 \epsilon}\diff W, 
\end{equation}
where $T$ is a decorrelation timescale, $C_0$ a universal constant and $\epsilon$ is the mean dissipation. Indeed, taking into account the definition of the particle position $x$, $ \diff x = u \diff t, $ we get a scaling of  the time averages of  velocity and  position  as:
\begin{equation}
  \overline{u^2(t)} \sim t, \quad   \overline{x^2(t)} \sim t^3.
\label{IOU}
  \end{equation}
The second property is the Richardson law. Then, defining $\delta u=  [\overline{u^2(t)}]^{1/2}$ and $\ell=[\overline{x^2(t)}]^{1/2}$, we get from eq.(\ref{IOU})  $\delta u \sim \ell^{1/3}$ which leads to the space refined Kolmogorov hypothesis and spectrum.\

The discrete time version of eq. (\ref{IOU})  can be written as:
\begin{equation}\label{AR1}
u_t = \phi u_{t-1} + \psi_t,
\end{equation}
where $t$ is a discrete time label, $\diff W$ are the increments of a Brownian motion, $ \phi=\left(1-\frac{\Delta t}{T}\right) $ and $\psi_t$ are independent variables, normally distributed. Eq. (\ref{AR1}) is the expression of an auto-regressive process of order one, noted AR$(1)$. To be able to account for  intermittency or memory effects that are present in real flows, it is then convenient to consider  a projection of the velocity data on higher order ARMA$(p,q)$ models
and define the intermittency as a distance with respect to the Kolmogorov AR(1) model in this space.\\

 \paragraph{Intermittency as a distance in ARMA space.}
A stationary time series $X_t$ is said to follow an ARMA$(p,q)$ process if it satisfies the discrete equation:
\begin{equation}
X_t = \sum_{i=1}^p \phi_i X_{t-i} + \varepsilon_{t} + \sum_{j=1}^q \theta_j \varepsilon_{t-j},
\label{ARMA1}
\end{equation}
with $\varepsilon_t  \sim WN(0, \sigma^2)$ - where $WN$ stands for white noise - and the polynomials $\phi(z) = 1 - \phi_1 z_{t-1} - \cdots - \phi_p z_{t-p}$ and $\theta(z) = 1 - \theta_1 z_{t-1} - \cdots - \theta_q z_{t-q}$, with $z \in \mathbb{C}$, have no common factors. Notice that the noise term $\varepsilon_t$ will be assumed to be a white noise, which is a very general condition \cite{brockwell_etal-springer-1990}.  We ensure unicity by  applying the  Box-Jenkis procedure \cite{box_etal-springer-1970}: we choose the lowest $p$ and $q$ such that the residuals of the series filtered by the process ARMA($p,q$) are not correlated. To define a suitable distance in the space of ARMA$(p,q)$ models, we introduce the Bayesian information criterion ($BIC$), measuring the relative quality of a statistical model, as: 
\begin{equation}\label{BIC}
BIC = -2 \ln{\hat{L}(n, \hat{\sigma}^2, p,q)} + k [\ln(n) + \ln(2 \pi)],
\end{equation}
where $\hat{L}(n, \hat{\sigma}^2, p,q)$ is the likelihood function for the investigated model and in our case $k=p+q$ and 
 $n$ the length of the sample. The   variance $\hat{\sigma}^2$ is computed from the sample and is a series-specific quantity. The normalized distance between the  fit ARMA$(p+1,q)$ and the Kolmogorov AR(1) model is then defined as the normalized difference between the  $BIC$($n, \hat{\sigma}^2, p+1,q$) and the AR(1)  $BIC$($n, \hat{\sigma}^2, 1,0$):
\begin{equation}\label{Intermittency}
\Upsilon= 1 - \exp \left \{| BIC(p+1,q) - BIC(1,0) | \right \}/n.
\end{equation}
with  $\quad 0\le \Upsilon\le 1$: it goes to zero if the dataset is well described by an AR(1) model and tends to one in the opposite case. When applied to time series of velocity increments, it is   a measure of the deviations from the Kolmogorov model. We introduce the $p+1$ correction to magnify small $\Upsilon$ values.\\

\begin{figure}
\includegraphics[width=.4\textwidth]{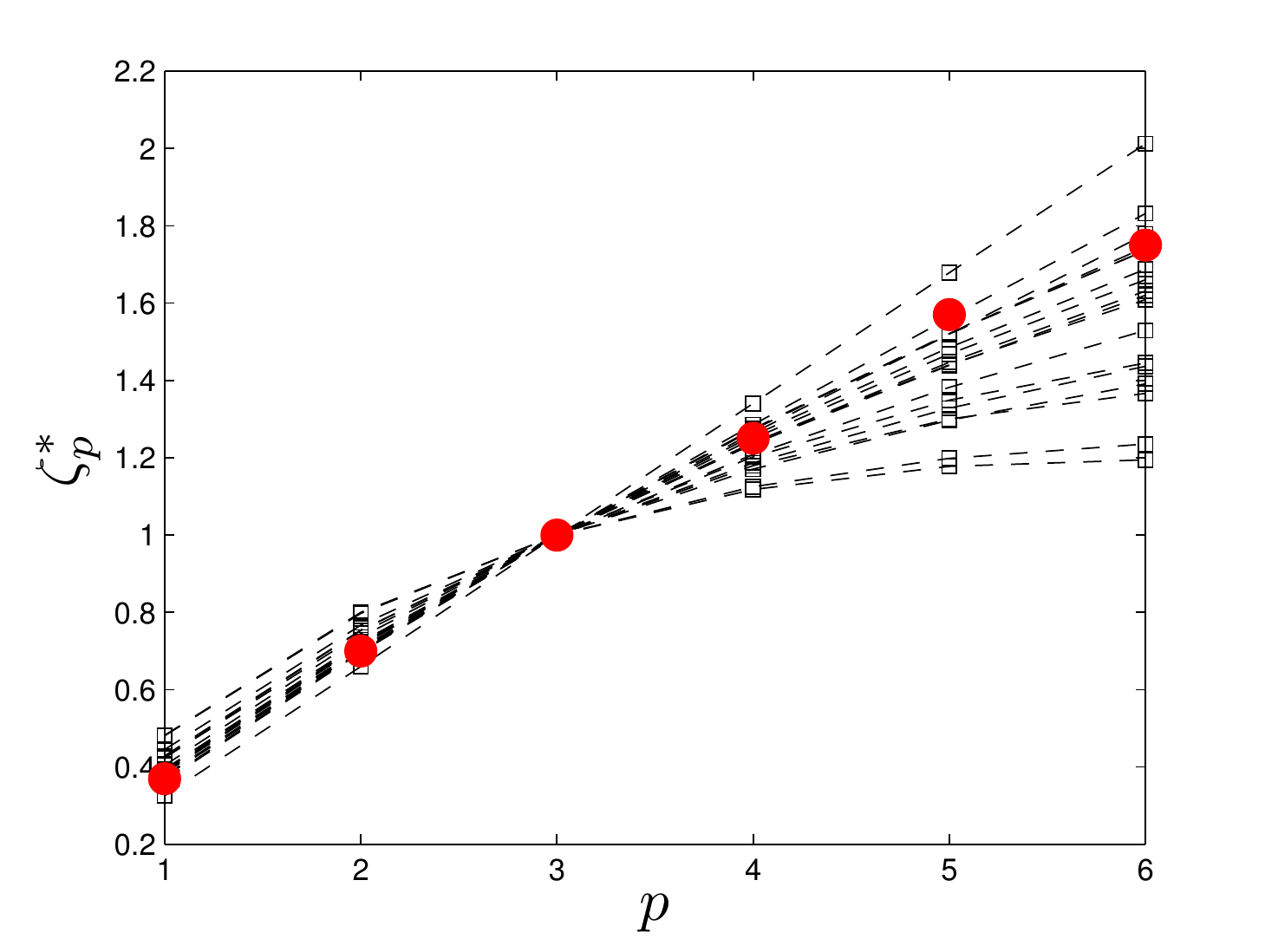}
\caption{$\zeta^*_p$ computed for the LDV experiments.  Different lines correspond to different measure points. Red spots mark the  scaling exponents reported in \cite{arneodo1996structure}.} 
\label{arne}
\end{figure}

\begin{figure}\includegraphics[width=.5\textwidth]{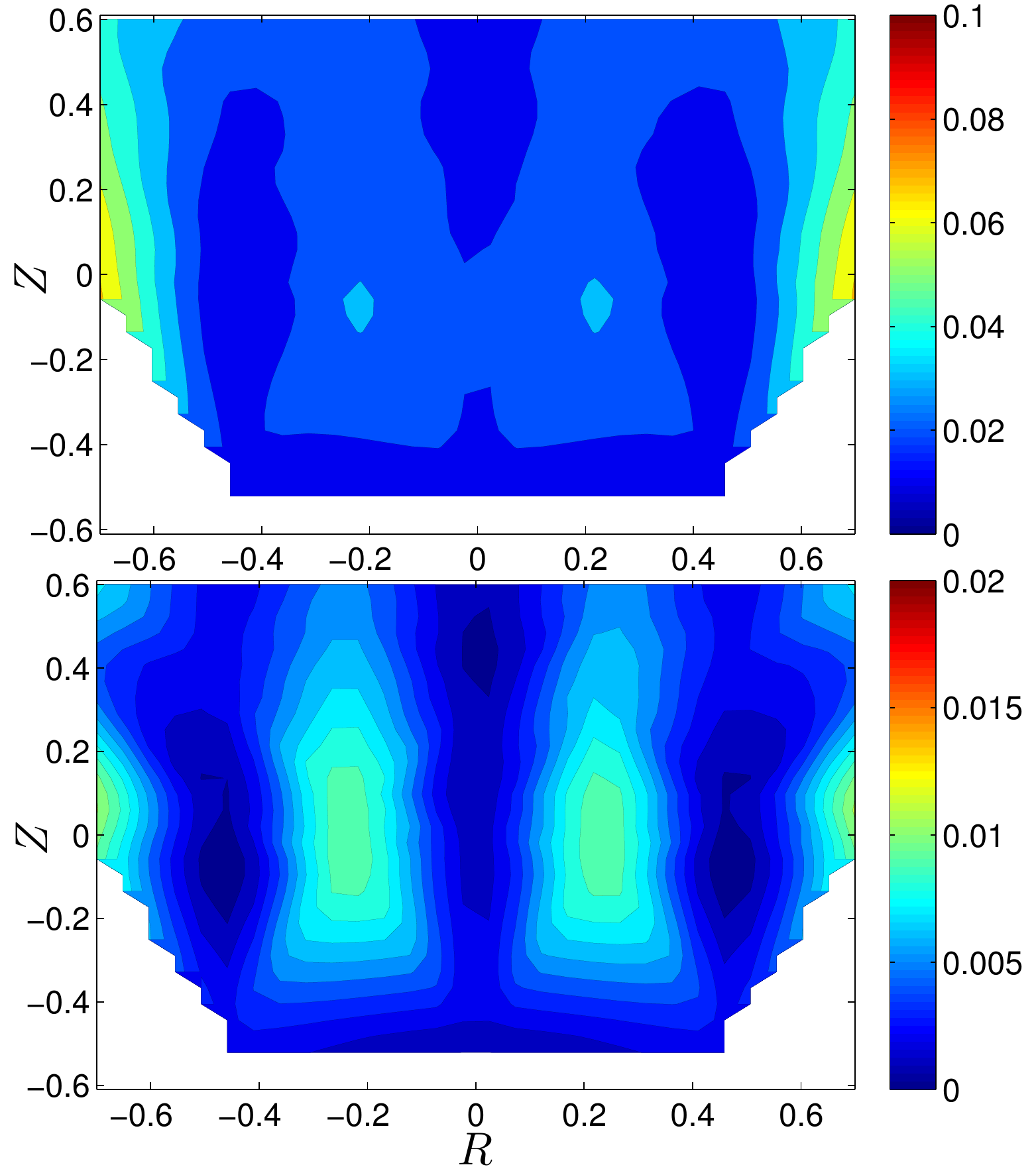}\\
\hskip -20pt
\includegraphics[width=.4\textwidth]{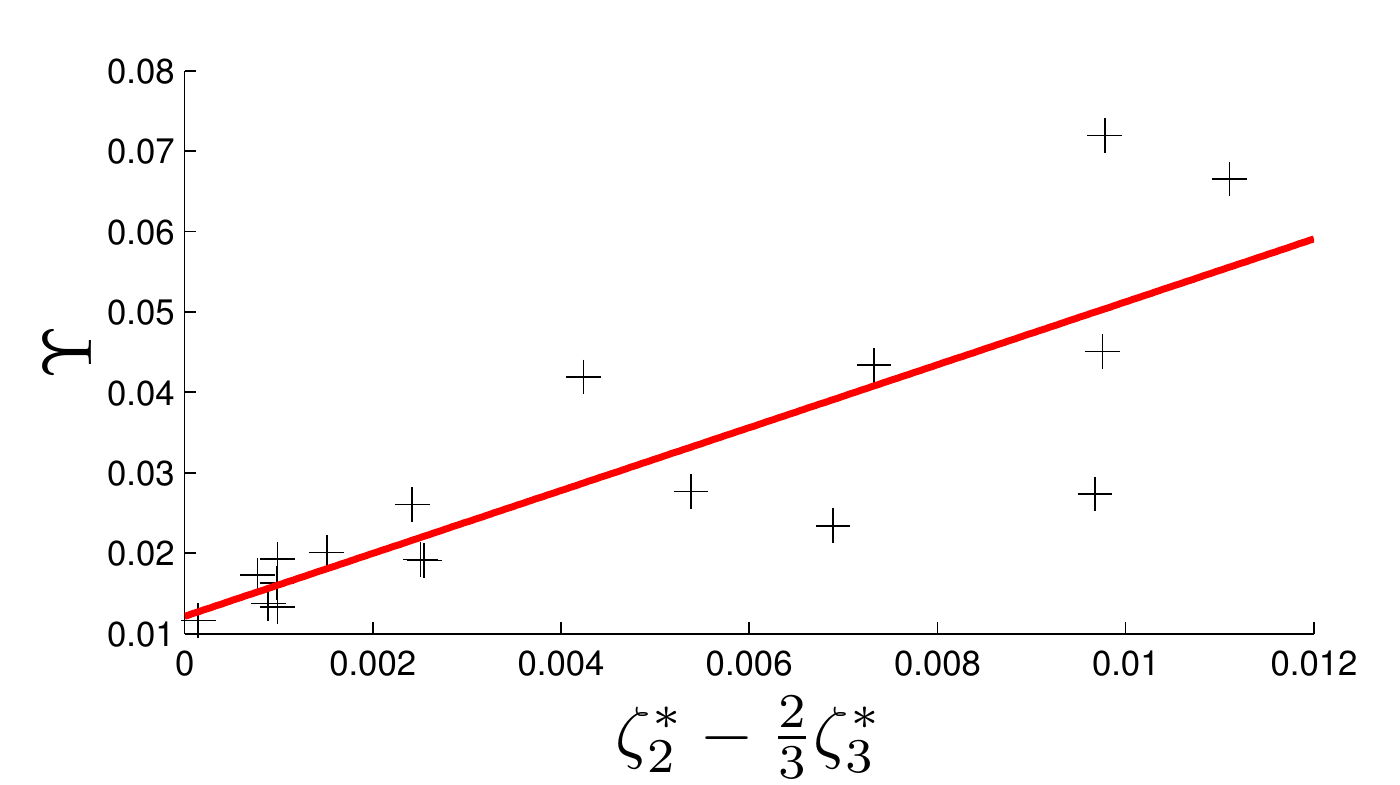}
\caption{Index $\Upsilon$ (upper panel) vs the intermittency index $\mu={\zeta^*_2}-\frac{2}{3}\zeta^*_3$ (central panel). Red crosses show measurement points. The lower panel shows a scatter plot of  $\Upsilon$  vs $\mu$ The red line shows a linear regression of the data.} 
\label{prop}
\end{figure}

\begin{figure}
\includegraphics[width=.50\textwidth]{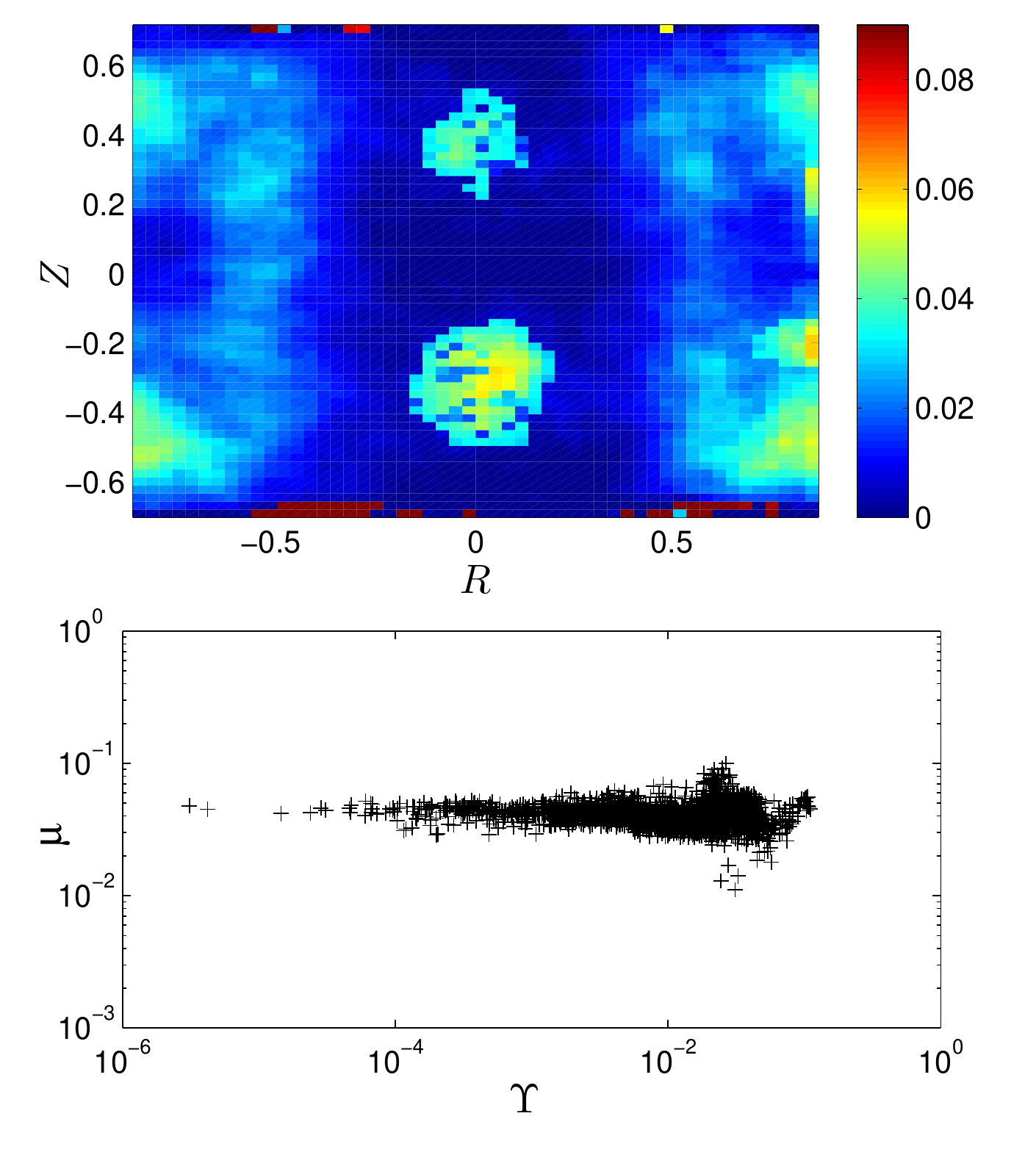}
\caption{Upper panel: $\Upsilon$ computed for a PIV experiment. The lower panel shows a scatter plot of  $\Upsilon$  vs $\mu$.} 
\label{upsi}
\end{figure}
\paragraph{Application to turbulent data.}
We apply the index defined in eq. \ref{Intermittency} to time series of velocity fields obtained in a  von K\'arm\'an turbulent swirling flow. The experimental set-up consists of two sets of blades mounted on two counter-rotating co-axial impellers at the top and bottom of a cylindric vessel of diameter $R=0.1$ m.  The operating fluid is water, the rotation frequency of the impellers can reach $F=15$ Hz, resulting in large   Reynolds numbers ($Re=2\pi F R^2\nu^{-1}  \sim 10^6$). A detailed description of  the experiment can be found in \cite{ravelet2008supercritical,cortet2010experimental,cortet2009normalized}. Two techniques are used to measure the fluid velocity on a grid: the Particle Interferometry Velocimetry (PIV) and the Laser Doppler Velocimetry (LDV),  mapped on a regular sampling time applying a sample-and-hold algorithm. The stereoscopic  PIV measures the three components of the velocity field into a plane, while the LDV measurements reported have given the out-of-plane velocity component $V_{\phi}$ into a plane. The PIV produces regularly sampled time series at intervals of $0.1$ s over a  sample size  at most of order $10^4$ and a spatial resolution of order of $1$ mm, \textit{i.e.} 10 to 100 times larger than the dissipation scale. The LDV time-series are sampled over  time-scale of the order of $0.001$ s, producing sample size  up to $10^6$ data on a grid of spatial resolution of the order of $1$ cm. Given these resolutions constraints, we  compute spatial (resp. temporal)  velocity increments for the PIV (resp. LDV) data. The idea is to compute at each spatial grid location the classical intermittency index   $\mu$, compare it to $\Upsilon$ , and see how they vary spatially.  All the analyses presented in this letter are done using the three components for the PIV and  $V_{\phi}$ for the LDV.
Since the von K\'arm\'an flow is  inhomogeneous and anisotropic with large fluctuations \cite{cortet2009normalized}, we expect that the time and space velocity structure functions depend on the measurement points. This is illustrated in Fig. \ref{g3l}, for the second and fourth order spatial and time structure functions. For the spatial case, deviations from the Kolmogorov scaling (solid lines) are small for the spatial structure functions, near the symmetry plane $Z=0$. This plane is the location of an intense shear layer, and has traditionally been used to perform "isotropic homogeneous" like measurements. Outside this plane,  deviations from the Kolmogorov scaling are large. For the time case, one observes two distinct behaviors: outside the shear layer, where a mean velocity is well-defined, one observes close to Eulerian Kolmogorov scaling at the smallest time increments ($\tau^{p/3}$). In the shear layer, where no Taylor hypothesis holds, the scaling is closer to Lagrangian scaling ($\tau^{p/2}$). However, as already noted by \cite{pinton} and shown in Fig. \ref{arne}, the {\sl relative} scaling exponents $\zeta^*_p$   computed as  $G_p(\tau)\sim \langle \vert\delta u_\tau\vert^3 \rangle^{\zeta^*_p}$ (ESS method) are in most of the flow close to the universal scaling exponents found by   \cite{arneodo1996structure}, in a variety of homogeneous turbulent flows even those with no obvious inertial range.   Using these ESS scaling exponents to compute the $\mu$ index, we may then draw a map of the intermittency and compare it with $\Upsilon$. This is done in Fig. \ref{prop} for an LDV experiment at $Re \sim 10^5$. The spatial patterns look indeed similar. Moreover, the plot of  $\Upsilon$  as a function of $\mu$ (lower panel of Fig. \ref{prop}) evidences a linear relation between them; the linear regression represented by the red line leads to a linear correlation coefficient $r \simeq 0.69$. This means that $\Upsilon$ traces the same intermittency characteristics as the time structure functions.  
The comparison of  $\Upsilon$, with the intermittency index $\mu$ computed for spatial structure functions is also informative: because of convergency issues, we have to use a data set of about $10^5$ to $10^6$ data points to converge the estimate of $\mu$, while only $10^3$ are  needed to converge $\Upsilon$. To illustrate this, we use the longest data set available,  a $9000$ velocity fields  of a PIV experiment  performed at $Re\simeq 5\cdot 10^4$. At this value, the von Karman flow experiences the equivalent of  a phase transition \cite{cortet2011susceptibility}, with time wandering of the shear layer in between $Z=0.3$ and $Z=-0.3$.  This corresponds to a very large time intermittency and  is clearly detected by the $\Upsilon$  index as shown in Fig. \ref{upsi}, under the shape of two patches at $R\simeq 0$,  $Z=0.3$ and $Z=-0.3$. This pattern is unique of the phase transition and is not present in other PIV experiments (see \cite{faranda2014modelling} for examples). Besides this, one observes a fairly symmetric structure, with maxima correspond to the four cells structure of the flow. However, the time intermittency prevents the convergency of the spatial structure functions, resulting in a lack of of symmetry of the $\mu$ field (not shown). As a result, $\mu$ fluctuates over a decade around a value of about 0.05,  while $\Upsilon$ spans several orders of magnitude, as can be seen in the lower panel of Fig. \ref{upsi}. This shows that  $\Upsilon$ is a more sensitive tool to detect intermittency than $\mu$.\\

\paragraph*{Discussion.} 
We have introduced an intermittency index $\Upsilon$ that can be intuitively interpreted as a statistical distance between the best fit linear model for a turbulent time series and the simplest possible process, i.e. an AR(1). To this purpose, we have exploited a Bayesian information criterion, opportunely normalized.
We have compared the obtained values of $\Upsilon$ with a classical  intermittency index $\mu=\zeta^*_2-\frac{2}{3}\zeta^*_3$; the two parameters show to be linearly related, with a coefficient of determination $R \simeq 0.69$ for the time structure function.
In the spatial case, while $\Upsilon$ clearly catches important characteristics of the mean flow, the lack of convergence of $\mu$ prevents us from a comparison. Therefore, the main advantage of this new index is the applicability to cases in which no big datasets are available: while a robust estimation of the structure functions requires very long time series, an ARMA($p,q$) process can be usually fitted even in case the series is of the order of $10^3$ data. In fact, as we have verified for the LDV data, the estimates of $\Upsilon$ do not change signifcantly when resampling the original series at a frequency comparable to the PIV one. This opens the possibility to compute $\Upsilon$  from  the much shorter PIV  time series, without requiring high-order moments computation.Our results reinforce the hypothesis firstly proposed by \cite{laval2001nonlocality} that intermittency \textit{propagates} in direct interactions between large and small scales, rather  than in cascades.\\
Moreover, several models of  complex systems are available in terms of a stochastic differential equation \cite{faranda2014statistical}. The methodology presented here  can be used to validate such models  with respect to experimental data.

\paragraph*{Acknowledgments.}   We acknowledge the other members of the VKE collaboration who performed the experiments: E. Herbert, P.F. Cortet, C. Wiertel and V. Padilla. DF acknowledges the support of a CNRS postdoctoral grant.

\bibliography{ARMAbib}

\end{document}